\documentclass[twocolumn,showpacs,amsmath,amssymb,pra]{revtex4}
\usepackage{graphicx}% Include figure files
\usepackage{dcolumn}% Align table columns on decimal point
\usepackage{bm}% bold math
\usepackage{subfigure}
\begin{document}

%\preprint{submitted to Phys. Rev. B}

\title {
Theory of elastic anomalies at the $\gamma \rightarrow \alpha$ phase transition in solid Ce
}

\author{K. H. Michel$^1$}

\author{A. V. Nikolaev$^{2,3}$}

\affiliation{$^1$Department of Physics, University of Antwerp,
Groenenborgerlaan 171, 2020, Antwerpen, Belgium}

\affiliation{$^2$Frumkin Institute of Physical Chemistry and
Electrochemistry of RAS, Leninskii pr. 31, 119071, Moscow, Russia}

\affiliation{$^3$Skobeltsyn Institute of Nuclear Physics, Moscow
State University, Vorob'evy Gory 1/2, 119991, Moscow, Russia}

\date{\today}

%--------------- ABSTRACT ---------------
\begin{abstract}
Starting from a model of 4f-electron generated quadrupolar densities on a compressible fcc lattice,
the elastic anomalies at the $Fm{\bar 3}m \rightarrow Pa{\bar 3}$ phase transition are studied by means
of analytical theory.
The model is taken as representative for the $\gamma-\alpha$ phase transition in Ce.
The coupling of the (linear) lattice displacements to the square of the quadrupolar orientational
density fluctuations renormalizes the elastic constants.
The condensation of the quadrupolar densities into the orientationally ordered $Pa{\bar 3}$ structure
is studied as function of temperature and pressure.
Precursor effects of the transition lead to an anomalous softening of the elastic constant $c_{11}$
while $c_{44}$ exhibits no such softening.
The theoretical results are in excellent qualitative agreement with pressure experiments on the elastic
constants (equivalently on sound velocities) at the $\gamma-\alpha$ transition in Ce.
Lattice dynamical analogies in theory and striking similarities in experimental results with the
$Fm{\bar 3}m \rightarrow Pa{\bar 3}$ transition in C$_{60}$ fullerite are discussed.
\end{abstract}


\maketitle

\section{Introduction}
\label{sec:int}

Cerium, the first of the lanthanide series (rare earths) with an inner 4f-electron exhibits
a broad range of unusual electronic, magnetic and structural properties in the solid state \cite{Kosk}.
Most intriguing is the transformation under pressure \cite{Bri} between two face-centered-cubic (fcc) phases
called $\gamma$-Ce and $\alpha$-Ce.
While $\gamma$-Ce is stable at room temperature ($T$) and ambient pressure ($P$), it transforms to $\alpha$-Ce
at $P \approx 0.8$ GPa.
This apparent isostructural phase transition $\gamma \rightarrow \alpha$ is accompanied by a $\sim$15{\%} volume
reduction and a change in magnetism.
While in the $\gamma$-phase the magnetic susceptibility $\chi(T)$ follows a Curie Weiss law which is attributed to
the magnetic moment of the 4f-electron, it is a quasi $T$-independent in the $\alpha$-phase and reminiscent of Pauli paramagnetism \cite{MPhe}.

The isostructural nature of the $\gamma \rightarrow \alpha$ transition is at variance with the Landau theory
of phase transitions \cite{Lan1} which would require a change of space group symmetry.
Since the establishing of the fcc structure for both phases \cite{Law} various microscopic theories have been proposed.
Nowadays two theoretical models are well represented in the literature.

A Mott-transition scenario for 4f-states is based on the concept of a localized non bonding state of the 4f-electron in
$\gamma$-Ce and an extended metallic bonding state in the $\alpha$-phase \cite{Joh}.
The increased binding capability due to the metallic f state is considered then as the reason for the marked volume reduction
in the $\alpha$-phase \cite{Joh}.
In the following this model has been supported by band structure calculations \cite{Joh2}.
On the other hand more recent inelastic neutron scattering experiments on $\alpha$-Ce found that the
magnetic form factor is quite different \cite{Mur} from the one calculated \cite{Hje} under the assumption that the 4f states are itinerant.
The absence of extended 4f states in the solid is also supported by many-electron calculations on the Ce dimer \cite{Cao,Roo,NikA}
where it is found that the 4f-electrons do not participate in the chemical bond.

The second widely recognized theoretical model is based on spin fluctuations.
These are the driving mechanism in the Kondo-volume-collapse (KVC) model of the $\gamma \rightarrow \alpha$ transition \cite{All,Lav}.
In first approximation the 4f-electron stays localized in both phases.
however in the $\alpha$-phase the hybridization between the 4f-electrons and the conduction electrons
(Anderson impurity Hamiltonian, see e.g. Ref. \onlinecite{Ful}) is much more intense than in the $\gamma$-phase.
This interaction leads to a screening of the 4f local moments in the energy ground state.
The volume dependence of the hybridization coupling or equivalently of the Kondo-temperature $T_K$ is used to
obtain the equation of state.
The interpretation of the $\gamma \rightarrow \alpha$ transition in Ce within the KVC scenario \cite{All} has found strong support by the analysis
of electron spectroscopy data \cite{Liu}.

Notwithstanding much effort, both the Mott transition and the KVC model remain under debate \cite{Joh1} and there is no consensus on the
driving mechanism of the phase transition \cite{Ama,Rue}.
A common feature is the absence of any symmetry breaking at the transition as would be required by Landau theory \cite{Lan1}.
On the basis of thermodynamical data it has been suggested \cite{EC} that the $\gamma \rightarrow \alpha$ transformation is in fact a
first order phase transition which becomes of second order beyond a tricritical point.
Therefore, $\alpha$-Ce should have lower symmetry than $\gamma$-Ce.
However a distorted lattice has been discarded by x-ray diffraction experiments \cite{Jeo,Lip}.

A mechanism of symmetry lowering without lattice distortion (positions of the Ce nuclei fcc in the $\gamma$ and $\alpha$ phase) has been suggested
by the present authors \cite{NM1a,NM1b,NM2}.
The main concepts are the support of quadrupolar electronic charge-density fluctuations by the 4f-electrons in the $\gamma$-phase on a compressible
fcc lattice and the collective orientational ordering of the quadrupolar densities in the $\alpha$ phase \cite{NM1a,NM1b,NM2} on four simple
cubic (sc) sublattices.
The space group symmetry lowering at the transition $\gamma \rightarrow \alpha$ is $Fm{\bar 3}m \rightarrow Pa{\bar 3}$,
the local density symmetry is $S_6$. It is accompanied by a uniform lattice contraction so that the fcc structure is conserved.
Although this latter aspect is ``isostructural", the symmetry lowering of the electronic structure is fully consistent with Landau theory.
In reciprocal space the active electronic density mode condenses at the $X$ point of the Brillouin zone.

A phase transition $Fm{\bar 3}m \rightarrow Pa{\bar 3}$ occurs in solid C$_{60}$ (fullerite) \cite{Sac,Dav}.
Here the symmetry lowering is due to the orientational ordering of the C$_{60}$ molecules.
At the transition the cubic lattice constant contracts discontinuously \cite{Dav2,Hei2}, the center of mass points of the molecules
still occupy an fcc lattice.
From the latter point of view, the phase transition would be ``isostructural".
However, the molecular order on four sc sublattices entails an order of the constituent C atoms and hence the $Pa{\bar 3}$ structure
leads to characteristic reflections in x-ray \cite{Sac} and neutron scattering \cite{Dav} experiments.

The problem of measuring the $Pa{\bar 3}$ structure directly in Ce is complicated by the fact that the Ce nuclei remain on an fcc lattice.
Although the quadrupolar charge density ordering could be detectable by synchrotron radiation, the intensity of the additional $Pa{\bar 3}$
reflections is likely to be too weak and has not yet been uncovered \cite{Dec}.
On the other hand there have been two recent experimental results which are very specific and point to the relevance of the
quadrupolar ordering at the $\gamma \rightarrow \alpha$ transition.

Time-differential perturbed angular correlation (TDPAC) experiments in solid Ce in a pressure range up to 8 GPa detect an appreciable
electric field gradient (EFG) in $\alpha$-Ce which is almost four times larger than in the cubic $\gamma$ phase and close to values in
the noncubic phases $\alpha'$ and $\alpha''$ \cite{Tsv}.
This finding rules out the $Fm{\bar 3}m$ symmetry in $\alpha$-Ce and evidences in support of the antiferroquadrupolar order suggested
by theory \cite{NM1a,NM1b,NM2}.

The other experimental support of the antiferroquadrupolar ordering is more indirect but very specific too.
Phonon dispersion measurements by inelastic x-ray scattering on elemental Ce across the $\gamma \rightarrow \alpha$ transition reveal
strong changes in the dispersion shape \cite{Kri}. In particular a pronounced softening of certain phonon branches is found in the
$\alpha$-phase toward the $X$ point of the Brillouin zone.

Given the above mentioned fact that the discussion on the validity of the Mott transition or KVC scenario remains open, there have been
in the last decade increased experimental efforts in investigating the lattice dynamics at the phase transition $\gamma \rightarrow \alpha$ \cite{Jeo,Lip,Dec1,Kri}.
From these studies it results that the lattice vibrations play an important role at the $\gamma \rightarrow \alpha$ phase transition
and that the inclusion of lattice dynamics in the theoretical description is of paramount importance.
In this respect we recall that anomalous elastic behavior at the $\gamma \rightarrow \alpha$ transition was originally discovered in
ultrasound experiments \cite{Vor}.
From inelastic neutron scattering experiments \cite{Sta} performed on single crystal of $\gamma$-Ce it was concluded that
premonitory effect of the $\gamma - \alpha$ transition are present in the phonon dispersion curves of $\gamma$-Ce at room temperature.
However, until recently the relevance of these results has not been sufficiently appreciated, while all research efforts were concentrated on
the electronic properties.

The anomalies of the elastic properties at the $\gamma \rightarrow \alpha$ transition in Ce [34,21,33] %\cite{Vor,Jeo,Dec1}
have their counterpart at the fcc$\rightarrow$sc transition in C$_{60}$ fullerite \cite{Schr,Lun}.
In particular there is a striking resemblance of the pressure dependence of the bulk modulus in Ce with corresponding
experimental results in C$_{60}$ \cite{Lun}.

The content of the paper is as follows.
In Sec.\ \ref{sec:model} we recall the main features of the model of interacting 4f-electron quadrupolar densities on a compressible fcc lattice.
The Hamiltonian and the resulting condensation scheme from the disordered $\gamma$-phase to the quadrupolar ordered $\alpha$-phase are described.
The coupling to the lattice is included. Next (Sec.\ \ref{sec:phase_trans}) we extend the model by adding an external hydrostatic pressure.
Various quantities that characterize the phase transition (order parameter, susceptibility and correlation length) are derived as
functions of temperature and pressure.
In Sec.\ \ref{sec:el_res} we calculate the dynamic displacement--displacement correlation function, taking into account
the coupling of the quadrupolar electron densities to the crystal lattice.
In the static limit the inversion of the displacement--displacement correlation function tensor leads to the elastic constants c$_{11}$ and $c_{44}$.
While in c$_{11}$ the coupling to quadrupolar electron density fluctuations leads to remarkable anomalies at the $\gamma-\alpha$ phase transition,
the coupling is absent in c$_{44}$.
The results of the theory are compared with experiments.
Conclusions of the paper are presented in Sec.\ \ref{sec:con}.

%%%%%%%%%%%%%%%%%%%%%%%%%%%%%%%%%%%%%%%%

% 2
\section{model}
\label{sec:model}

Here we will remind the model Hamiltonian which describes electronic quadrupolar charge density fluctuations on a compressible fcc lattice.
In previous work we have first treated quadrupolar charge fluctuations due to solely 4f-electrons \cite{NM1a}.
Later on we have taken into account conduction electrons \cite{NM1b} as well as f- and d-electron intra site correlations \cite{NM2}.
Since these extensions can be cast into an effective Hamiltonian with essentially the same structure as the one originally studied \cite{NM1a},
we will restrict ourselves here to 4f-electrons as a generic case.

We consider $N$ Ce atoms located on a non rigid fcc lattice with nuclear positions
$\vec{R}(\vec{n})=\vec{X}(\vec{n})+\vec{u}(\vec{n})$, $\vec{n}=1$, 2, ... , $N$. Here $\{ \vec{X}(\vec{n}) \}$ denote the equilibrium positions
and $\{ \vec{u}(\vec{n}) \}$ the displacements due to lattice vibrations.
We assume that core and valence electrons follow adiabatically the nuclear displacements.
The Hamiltonian comprises three parts:
\begin{eqnarray}
     H = H^{(e)} + H^{(L)} + H^{(e,L)} ,
\label{m1}
\end{eqnarray}
where $H^{(e)}$ stands for the 4f-electrons, $H^{(L)}$ for the lattice and $H^{(e,L)}$ for the coupling of both.

Explicitly one has
\begin{eqnarray}
     H^{(e)} = U_0 + U_{QQ} ,
\label{m2}
\end{eqnarray}
where $U_0$ is the single particle potential and $U_{QQ}$ the quadrupole-quadrupole interaction on the rigid lattice.
We assume that the 4f-electrons have coordinates $\{ \vec{X}(\vec{n})+\vec{r}(\vec{n}) \}$ and
are localized on spheres centered at $\{ \vec{X}(\vec{n}) \}$ and with radii $\{ \vec{r}(\vec{n}) \}$,
$|\vec{r}(\vec{n})|=r_f=$1.378 a.u. \cite{NM1b}.
The single particle potential is due to the cubic crystal field in presence of spin-orbit coupling.
One obtains
\begin{eqnarray}
     U_0 = \sum_{\vec{n}} \sum_i | i \rangle_{\vec{n}} \epsilon_i \langle i |_{\vec{n}} ,
\label{m3}
\end{eqnarray}
where $\epsilon_i$ and $| i \rangle_{\vec{n}}$ are the eigenvalues and eigenstates, $i=1$, 2, ... , 14.
The angular part of the wave function of the 4f-electron (angular momentum quantum number $l=3$)
at site $\vec{n}$ is described by the $2(2l+1)$ spin orbitals
$\langle i | \hat{n} \rangle$, $\hat{n} \equiv \Theta(\vec{n})$, where $\Theta = (\Theta, \phi)$.
For details see Refs.\ \onlinecite{NM1a,NM1b,NM2}.
The electronic quadrupolar interaction is given by
\begin{eqnarray}
  U_{QQ} = {\frac{1}{2}} {\sum_{\vec{n}, \vec{n}'} }' \sum_{\alpha \beta}
  \rho_{Q}^{(\alpha)}(\vec{n})\, v_{\alpha \beta}(\vec{n}-\vec{n}')\,
  \rho_{Q}^{(\beta)}(\vec{n}') ,
\label{m4}
\end{eqnarray}
where $\rho_{Q}^{(\alpha)}(\vec{n})$, $\alpha(\beta)=$1, 2, 3 are quadrupolar density operators and $v_{\alpha, \beta}(\vec{n}-\vec{n}')$
interaction matrix elements between nearest neighbors on the fcc lattice.
The density operators are defined by
\begin{eqnarray}
  \rho_{Q}^{(\alpha)}(\vec{n}) = \sum_{ij} c_Q^{(\alpha)}(ij) | i \rangle_{\vec{n}} \langle i |_{\vec{n}} ,
\label{m5}
\end{eqnarray}
with
\begin{eqnarray}
   c_Q^{(\alpha)}(ij) = \int d\Omega \, \langle i | \hat{n} \rangle S^{(\alpha)}(\hat{n})\, \langle \hat{n} | j \rangle .
\label{m6}
\end{eqnarray}
Here we restrict ourselves to the three quadrupolar functions $S^{(\alpha)}(\hat{n})$, $\alpha=1$, 2, 3, which transform
as the irreducible representation $T_{2g}$ of the cubic point group $m3m$ ($O_h$).
These symmetry adapted functions (SAF's) are linear combinations of spherical harmonics
belonging to the manifold $l=2$ and are tabulated in Ref.\ \onlinecite{BC}.
They transform as the Cartesian components $yz$, $zx$, $xy$ for $\alpha=1$, 2, 3, respectively.
Defining Fourier transforms
\begin{eqnarray}
   \rho_Q^{(\alpha)}(\vec{q}) = {\frac{1}{\sqrt{N}}} \sum_{\vec{n}}
   e^{-i\vec{q} \cdot \vec{X}(\vec{n})} \rho_Q^{(\alpha)}(\vec{n}) ,
\label{m7}
\end{eqnarray}
\begin{eqnarray}
  v_{\alpha \beta}(\vec{q}) = {\sum_{\vec{n}'}}'
  e^{-i\vec{q} \cdot (\vec{X}(\vec{n}')-\vec{X}(\vec{n}))}
  v_{\alpha \beta}(\vec{n}-\vec{n}') ,
\label{m8}
\end{eqnarray}
we obtain
\begin{eqnarray}
 U_{QQ}={\frac{1}{2}} \sum_{\vec{q}} \sum_{\alpha \beta}
  \rho_Q^{(\alpha)}(\vec{q})\, v_{\alpha \beta}(\vec{q})\,
  \rho_Q^{\beta}(-\vec{q}) ,
\label{m9}
\end{eqnarray}
with \cite{NM2}
\begin{eqnarray}
 & & v(\vec{q}) =  \nonumber \\
 & & {\scriptsize
 4 \left[
 \begin{array}{@{\hspace{-0.5mm}}c@{\hspace{-4mm}}c@{\hspace{-4mm}}c@{\hspace{-0.5mm}}}
 \gamma C_{yz} + \alpha (C_{zx}\!+\!C_{xy}\!) & -\beta S_{xy} & -\beta S_{zx} \\
 -\beta S_{xy} & \gamma C_{zx} + \alpha (C_{xy}\!+\!C_{yz}\!) & -\beta S_{yz} \\
 -\beta S_{zx} & -\beta S_{yz} &
 \gamma C_{xy} + \alpha (C_{yz}\!+\!C_{zx}\!)
\end{array} \right]
}
  \nonumber \\
  \label{m10}
\end{eqnarray}
Here $\gamma$, $\beta$ and $\alpha$ are quadrupole-quadrupole interaction coefficients on the fcc lattice.
The quantities $C_{ij}=\cos(q_i a/2) \cos(q_j a/2)$, and $S_{ij}=\sin(q_i a/2) \sin(q_j a/2)$, $i=x,y,z$,
$a$ is the cubic lattice constant, account for the fcc lattice structure.
At the $X$ point of the Brillouin zone for $\Gamma_c^f$, the interaction matrix $v(\vec{q})$ becomes
diagonal. The star of $\vec{q}^X$ has the arms $\vec{q}_x^X=(2\pi/a)(1,0,0)$, $\vec{q}_y^X=(2\pi/a)(0,1,0)$,
$\vec{q}_z^X=(2\pi/a)(0,0,1)$. For $\vec{q}=\vec{q}_x^X$ one has $v_{2 2}(\vec{q}_x^X)=v_{3 3}(\vec{q}_x^X)=-4\gamma$,
and similarly for $\vec{q}_y^X$ and $\vec{q}_z^X$ with permutation of indices $(\alpha \alpha)$.
Hence the quadrupolar interaction becomes attractive with the largest twofold degenerate eigenvalue
$\lambda_{X_5^+}=-4\gamma$ at each arm of $\vec{q}_x^X$.

\begin{figure}
\resizebox{0.35\textwidth}{!} {
\includegraphics{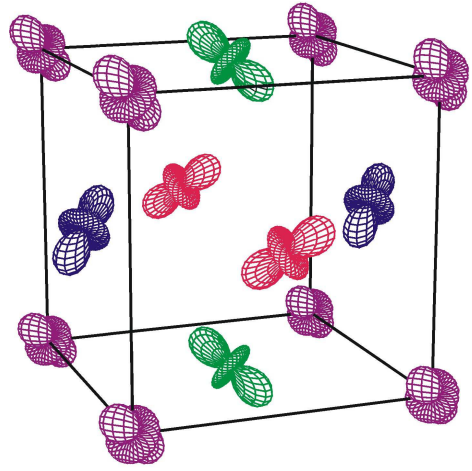}
}

\caption{ Triple-$\vec{q}$ antiferroquadrupolar ($Pa{\bar 3}$)
structure of $\alpha$-Ce proposed earlier
\cite{NM1a}. Quadrupoles represent the $l = 2$ valence electron
($4f+5d6s^2$) charge density distribution.
}
\label{fig0}

\end{figure}

As we have shown previously \cite{NM1a,NM1b,NM2} this interaction is compatible with a symmetry lowering
$Fm{\bar 3}m \rightarrow Pa{\bar 3}$, characterized by an orientational order of the quadrupolar
densities on four different sc sublattices (see Fig.\ \ref{fig0}).
%%%%%%%%%%%%%%%%%%%%%%%%%%%%%%%%%%%%%%%%%%%%%%%%%%%%%%%%%%%%%%%%%%
%
%------------------------------------------------------
%    FIGURE 1
%------------------------------------------------------
%
%%%%%%%%%%%%%%%%%%%%%%%%%%%%%%%%%%%%%%%%%%%%%%%%%%%%%%%%%%%%%%%%%
The corresponding condensation scheme reads:
\begin{subequations}
\begin{eqnarray}
 \rho_Q^{(3)e}(\vec{q}_x^X) = \rho_Q^{(1)e}(\vec{q}_y^X)
 = \rho_Q^{(2)e}(\vec{q}_z^X) \equiv \rho \sqrt{N} \neq 0, \quad
  \label{m11a} \\
 \rho_Q^{(2)e}(\vec{q}_x^X) = \rho_Q^{(3)e}(\vec{q}_y^X)
 = \rho_Q^{(1)e}(\vec{q}_z^X) = 0 . \quad \quad
  \label{m11b}
\end{eqnarray}
\end{subequations}
Here the superscript $e$ stands for a thermal expectation value and $\rho$ is the
order parameter amplitude.
Since the three arms of $\vec{q}_x^X$ are involved one speaks of a triple-$\vec{q}$
antiferroquadrupolar order (AFQ).
Notice that a condensation scheme similar to Eqs.\ (\ref{m11a}), (\ref{m11b}) also holds for
the phase transitions with the symmetry $Fm{\bar 3}m \rightarrow Pa{\bar 3}$ in NaO$_2$ \cite{Zie} and in C$_{60}$-fullerite \cite{Mic1}.
Although icosahedral symmetry of the C$_{60}$ molecule implies that not quadrupoles but higher order multipoles
($l=6$, 10, ...) determine the orientational interactions, the corresponding SAF's transform as irreducible
representations of symmetry $T_{2g}$ of the cubic point group.

A compelling mathematical reason to consider the symmetry lowering $Fm{\bar 3}m \rightarrow Pa{\bar 3}$
as a candidate for the isostructural phase transition in Ce is the fact that it leads to a lattice contraction
while the center of mass points still occupy an fcc lattice.
In addition the transition is of first order. Indeed the number of symmetry elements is thereby reduced
by a factor 3 (from 48 to 16), which implies the existence of a third order cubic invariant in free energy \cite{Lan1}.

The lattice dynamics is described by the phonon Hamiltonian
\begin{eqnarray}
     H^{(L)} = K + U ,
\label{m12}
\end{eqnarray}
where $K$ is the kinetic energy and $U$ the potential energy in harmonic approximation.
In Fourier space one has
\begin{subequations}
\begin{eqnarray}
 & & K = \frac{1}{2} \sum_{\vec{q}} \sum_i \vec{p}_i^{\dagger}(\vec{q}) \vec{p}_i(\vec{q}) ,
  \label{m13a} \\
 & & U = \frac{1}{2} \sum_{\vec{q}} \sum_{ij} M_{ij}(\vec{q}) u_i^{\dagger}(\vec{q}) u_j(\vec{q}) ,
  \label{m13b}
\end{eqnarray}
\end{subequations}
$i(j)=x,y,z$. The displacements $\vec{u}_i(\vec{q})$ and the conjugate moments $\vec{p}_i(\vec{q})$ are
related to the variables in real space by
\begin{subequations}
\begin{eqnarray}
   & & u_i(\vec{n}) = {\frac{1}{\sqrt{Nm}}} \sum_{\vec{q}} u_i(\vec{q})
   e^{i\vec{q} \cdot \vec{X}(\vec{n})} ,
\label{m14a}
\end{eqnarray}
\begin{eqnarray}
   & & p_i(\vec{n}) = {\sqrt{\frac{N}{m}}} \sum_{\vec{q}} p_i(\vec{q})
   e^{i\vec{q} \cdot \vec{X}(\vec{n})} ,
\label{m14b}
\end{eqnarray}
\end{subequations}
where $m$ is the atomic mass. One has the usual commutation rules $[u,u]=[p,p]=0$ and
\begin{eqnarray}
    [u_i(\vec{q}), p_j^{\dagger}(\vec{k})] = i \hbar \delta_{\vec{q} \vec{k}} \delta_{ij} .
\label{m15}
\end{eqnarray}
In the long wavelength limit the dynamical matrix $M(\vec{q})$ is given by
\begin{eqnarray}
 & & M(\vec{q}) = \frac{a^3}{4m}  \nonumber \\
 & & {\scriptsize
 \times \left[
 \begin{array}{ c c c }
 q_x^2 c_{11}^0 + (q_z^2+q_y^2)c_{44}^0  &  q_x q_y (c_{12}^0+c_{44}^0)  &  q_x q_z (c_{12}^0+c_{44}^0) \\
 q_y q_x (c_{12}^0+c_{44}^0)  &  q_y^2 c_{11}^0 + (q_x^2+q_z^2)c_{44}^0  &  q_y q_z (c_{12}^0+c_{44}^0) \\
 q_z q_x (c_{12}^0+c_{44}^0) &  q_z q_y (c_{12}^0+c_{44}^0)  &   q_z^2 c_{11}^0 + (q_x^2+q_y^2)c_{44}^0
\end{array} \right]
}
  \nonumber \\
  \label{m16}
\end{eqnarray}
where $c_{ij}^0$ are the bare elastic constants in absence of coupling to the quadrupolar electronic fluctuations.

The coupling between quadrupolar charge density fluctuations and lattice dynamics has been derived previously \cite{NM1a}.
In the long wavelength regime for the lattice displacements and for quadrupolar fluctuations near the $X$ point
of BZ we have
\begin{eqnarray}
    H^{(e,L)} = \frac{i}{2} \sum_{\vec{q}} \sum_{\vec{p}} \sum_{i} {\sum_{\alpha} }'
    {v_{i,\, \alpha \alpha} }'(\vec{q},\vec{p}) \nonumber \\
    u_i(\vec{q}) \, \rho_Q^{(\alpha)}(-\vec{p}-\vec{q}) \rho_Q^{(\alpha)}(\vec{p}) .
\label{m17}
\end{eqnarray}
Here the prime on the sum over $\alpha$ indicates the following restrictions:
for $i=x$, $\alpha=2,3$; $i=y$, $\alpha=3,1$; $i=z$, $\alpha=1,2$.
The coupling matrix is given by
\begin{eqnarray}
 v'_{i,\,\alpha \alpha}(\vec{q},\vec{p})= \frac{1}{\sqrt{Nm}} \sum_{\vec{h}} (\vec{q} \cdot \vec{X}(\vec{h}))
 v'_{i,\,\alpha \alpha}(\vec{h}) \cos(\vec{p} \cdot \vec{X}(\vec{h})) . \nonumber \\
 \label{m18}
\end{eqnarray}
Here $\vec{X}(\vec{h}) = \vec{X}(\vec{n}') - \vec{X}(\vec{n})$, $\vec{n}'$ refers to the 12 nearest neighbors
of $\vec{n}$ on the fcc lattice, $v'_{i,\,\alpha \alpha}(\vec{n}' - \vec{n})$ is the first order derivative
with respect to the lattice displacement component $i$ of the quadrupole-quadrupole interaction $v_{\alpha \alpha}(\vec{n}' - \vec{n})$.
The structure of the fcc lattice implies that $v'_{i,\,3 3}(\vec{h}_1)=v'_{i,\,2 2}(\vec{h}_2)=v'_{i,\,1 1}(\vec{h}_3) \equiv \Lambda$,
where for $i=x$ or $y$, $\vec{X}(\vec{h}_1)=a/2(1,1,0)$; for $j=z$ or $x$, $\vec{X}(\vec{h}_2)=a/2(1,0,1)$;
for $i=z$ or $y$, $\vec{X}(\vec{h}_3)=a/2(0,1,1)$.
Within the quadrupolar model we obtain that $\Lambda < 0$.
Carrying out the summation over $\vec{h}$ and exploiting the symmetry of the lattice, we rewrite Eq.\ (\ref{m17}) as
\begin{eqnarray}
    H^{(e,L)} = -\frac{i\Lambda a}{\sqrt{Nm}} \sum_{\vec{q}} \sum_{\vec{p}} \sum_{i} {\sum_{\alpha} }'
     q_i u_i(\vec{q})\, \nonumber \\
     \rho_Q^{(\alpha)}(-\vec{p}-\vec{q}) \rho_Q^{(\alpha)}(\vec{p}) .
\label{m19}
\end{eqnarray}
Under the proviso of the summation restriction over $\alpha$ we define for $i=x,y$ or $z$
\begin{eqnarray}
    \Omega^{(i)}(\vec{q}) = \sum_{\vec{p}} {\sum_{\alpha} }'
     \rho_Q^{(\alpha)}(-\vec{p}+\vec{q}) \rho_Q^{(\alpha)}(\vec{p}) ,
\label{m20}
\end{eqnarray}
and hence
\begin{eqnarray}
    H^{(e,L)} = -\frac{i\Lambda a}{\sqrt{Nm}} \sum_{\vec{q}} \sum_{i}
     q_i u_i(\vec{q})\, \Omega^{(i)}(-\vec{q}) .
\label{m21}
\end{eqnarray}
Only longitudinal lattice displacements or equivalently longitudinal lattice strains $\epsilon_{ii}$
occur on the right hand side of Eq.\ (\ref{m19}).
Since in addition $q_i u_i$ for $i=x,y$ and $z$ occur on the same footing, the coupling $H^{(e,L)}$ leads to a striction of
the lattice with conservation of cubic symmetry \cite{NM1a}.
Notice that there is no coupling to transverse lattice displacement waves or equivalently to shear strains $\epsilon_{ij}$,
$i \neq j$, in $H^{(e,L)}$.
In the next section we will investigate the influence of the quadrupolar density fluctuations on the low frequency
lattice dynamics.

%%%%%%%%%%%%%%%%%%%%%%%%%%%%%%%%%%%%%%%%%%%%%%%%%%%%%%%%%%%%%%%%%%%%%%%%%%%%%%%%%%%%
% 3
\section{Phase Transition}
\label{sec:phase_trans}

In order to study the tripple-$\vec{q}$ antiferroquadrupolar phase transition on a compressible lattice
as a function of temperature and pressure, we first recall some concepts of the underlying free energy.
The phase transition is of first order. We calculate the order parameter as a function of temperature and pressure.
We show that the transition temperature increases linearly with pressure.
Finally we study the order parameter susceptibility in the disordered and the ordered phase.

Taking into account the quadrupole-quadrupole interaction on a rigid fcc lattice we have written
the Helmholtz free energy as a Landau expansion in terms of the order parameter amplitude $\rho$ \cite{NM1a}:
\begin{eqnarray}
 F_{QQ}/N = F_0/N + A\rho^2+B\rho^3+C\rho^4 .
\label{fe22}
\end{eqnarray}
Here $F_0$ is the free energy of the disordered phase which has to be calculated with the cubic crystal field $U_0$,
Eq.~(\ref{m3}).
The coefficients of the order parameter terms are
\begin{eqnarray}
 A &=& {\frac{3}{2}}\,\left[\,{\frac{T}{x^{(2)}}} - 4\gamma \right], \label{fe23}   \\
 B &=& -T\,x_{123}^{(3)}\,/[x^{(2)}]^3 , \label{fe24} \\
 C &=& {\frac{T}{8(x^{(2)})^4}}\,
 \left[ \,9(x^{(2)})^2-x_{1111}^{(4)}-6x_{1122}^{(4)}+
 {\frac{24(x_{123}^{(3)})^2}{x^{(2)}}} \right] , \nonumber \\
\label{fe25}
\end{eqnarray}
where $4 \gamma$ is the quadrupolar interaction and $T$ the temperature in unergy units ($k_B=1$).
The quantities $x^{(2)}$, $x^{(3)}$ and $x^{(4)}$ are single particle thermal expectation values \cite{NM1a},
which are calculated by means of the single particle potential.
Numerical values of relevant parameters are given in Table 1.
%%%%%%%%%%%%%%%%%%%%%%%%%%%%%%%%%%%%%%%%%%%%%%%%%%%%%%%
%    TABLE 1
%
\begin{table}[!t]
\caption{
 Parameters $x^{(2)}$, $x^{(3)}$, $\gamma$, $\Lambda$, $B$, $C$, $C'$, $D$ calculated from Ref.\ \onlinecite{NM1a}. $P_1$ and
 $\kappa_L$ estimated from experiment \cite{Kri}.
\label{table1}     }
\begin{ruledtabular}
 \begin{tabular}{ c c }
  $x^{(2)}=23.34 \times 10^{-3}$   &   $x^{(3)}=-1.30 \times 10^{-3}$  \\
  $4 \gamma=3491$ K                & $\Lambda=445$ K/{\AA}  \\
  $B=8813$ K                       & $C=103964$ K  \\
  $C'=103692$ K                    & $D=216700$ K \\
  $a_{\gamma}=5.169$ {\AA}         & $a_{\alpha}=4.857$ {\AA}  \\
  $\kappa_L=(37.0$ GPa$)^{-1}$     & $P_1=0.8$ GPa \\

 \end{tabular}
\end{ruledtabular}
\end{table}
%%%%%%%%%%%%%%%%%%%%%%%%%%%%%%%%%%%%%%%%%%%%%%%%%%%%%%%

In case of a compressible lattice \cite{Lam3} the Hamiltonian $H^{(e,L)}$, Eq.\ (\ref{m21}),
gives the free energy term
\begin{eqnarray}
 F_{QQT}/N = -2a\Lambda \rho^2 [\epsilon_{xx} + \epsilon_{yy} + \epsilon_{zz}] ,
\label{fe28}
\end{eqnarray}
where $\epsilon_{xx}$ etc. are the lattice strains.
Since $\Lambda < 0$, this term favors $\epsilon_{ii} < 0$, i.e. a volume contraction.
The lattice Hamiltonian $H^{(L)}$ leads to the elastic contribution
\begin{eqnarray}
 F_{TT}/N &=& \frac{V_c}{2} [c_{11}^0(\epsilon_{xx}^2 + \epsilon_{yy}^2 + \epsilon_{zz}^2) \nonumber \\
        & & + 2c_{12}^0(\epsilon_{xx} \epsilon_{yy} + \epsilon_{yy} \epsilon_{zz} + \epsilon_{zz} \epsilon_{xx}) \nonumber \\
        & & + 4c_{44}^0(\epsilon_{xy}^2 + \epsilon_{yz}^2 + \epsilon_{zx}^2)] ,
\label{fe29}
\end{eqnarray}
where $V_c=a^3/4$ is the volume per atom.

In presence of an applied external pressure $P$, we consider the Gibbs free energy \cite{Lam3}
\begin{eqnarray}
 G = F_{QQ} + F_{QQT} + F_{TT} + PV(\epsilon_{xx} + \epsilon_{yy} + \epsilon_{zz}) .
\label{fe30}
\end{eqnarray}
Minimizing $G$ with respect to $\epsilon_{ii}$ for $i=x,y$ and $z$, we get an isostructural contraction
of the cubic lattice
\begin{eqnarray}
 \epsilon_{xx} = \epsilon_{yy} = \epsilon_{zz} = -[8a^{-2} |\Lambda| \rho^2 + P] \kappa_L .    % (29)
\label{fee29}
\end{eqnarray}
Here $\kappa_L=(c_{11}^0+2c_{12}^0)^{-1}$ is the compressibility.
Isostructural contraction of the lattice means that there is no symmetry breaking associated with
the change of the center of mass positions of the Ce atoms at the $\gamma \rightarrow \alpha$ transition.
We completely agree with conclusions from high-pressure and high-temperature x-ray diffraction experiments
that the structure remains fcc across the $\gamma \rightarrow \alpha$ transformation and retains
crystallographic orientation during the transformation \cite{Moo}.
Within the present theory a symmetry change occurs solely in the quadrupolar electronic charge densities
whereby the $Pa{\bar 3}$ ordering of the latter is compatible with the fcc structure of the lattice.

Eliminating $\epsilon_{ii}$ from Eq.\ (\ref{fe30}) and retaining only linear terms in $P$
we rewrite $G$ as
\begin{eqnarray}
 G[\rho]/N = F_0/N + A' \rho^2 + B \rho^3 + C'\rho^4 , \nonumber \\          % (30)
\label{fe31}
\end{eqnarray}
where
\begin{eqnarray}
  A' = \frac{3}{2} \left[ \frac{T}{x^{(2)}} - 4 \tilde{\gamma} \right] ,    % (31)
\label{fe32a}
\end{eqnarray}
with
\begin{eqnarray}
  \tilde{\gamma} = \gamma + a |\Lambda| \kappa_L P ,    % (32)
\label{fe32b}
\end{eqnarray}
and where
\begin{eqnarray}
 C' = C - 24 a^{-1} \Lambda^2 \kappa_L ,         % (33)
\label{fe32}
\end{eqnarray}
with $C'>0$.
The occurrence of the cubic invariant in the free energy implies that the phase
transition is of first order.
The coexistence condition $B^2-4A'C'=0$ establishes a relation between
the transition temperature $T_1$ and the corresponding pressure $P_1$:
\begin{eqnarray}
 T_1=x^{(2)}(T_1)
 [4 \gamma + {\frac{B^2(T_1)}{6C'(T_1)}} + 4 a |\Lambda| \kappa_L P_1 ] .    % (34)
\label{fe33}
\end{eqnarray}
In first approximation $T_1$ increases linearly with pressure, in agreement with the experimental
$(T,P)$ phase diagram of Ce \cite{Kosk}.
Notice that the analogue of Eq.\ (\ref{fe33}) also holds for C$_{60}$ fullerite \cite{Lam3},
where experiments \cite{Sam,Kriz} show a linear pressure dependence of $T_1$.

\begin{figure}
\resizebox{0.4\textwidth}{!} {
\includegraphics{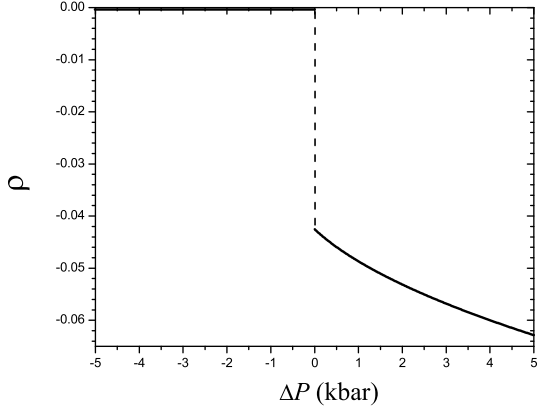}
}

\caption{ Pressure dependence of the order parameter amplitude $\rho$, $P_1$ taken as origin.
}
\label{fig1}

\end{figure}

With the parameters of Table \ref{table1} we obtain $T_1=89$ K.
In accordance with the experimental situation \cite{Jeo,Lip,Dec1,Vor}
we consider a fixed temperature $T=T_1$ and a variable pressure.
Then $P^< = P_1 + \triangle P$, $\triangle P \le 0$, corresponds to the disordered phase
(in casu $\gamma$-Ce), while $P^> = P_1 + \triangle P$, $\triangle P > 0$
corresponds to the ordered phase ($\alpha$-Ce).
Minimization of $G$ at $T_1$ and $P=P^>$ with respect to $\rho$ leads to the
order parameter amplitude
\begin{eqnarray}
 \rho(T_1,P^>) = \frac{-3B(T_1) - \sqrt{9B^2(T_1)-32A'(T_1,P^>)\,C'(T_1)}}{8C'(T_1)} . \nonumber \\   %(35)
\label{fe33b}
\end{eqnarray}
Since $B > 0$, the order parameter $\rho$ has to be negative and hence the $-$ sign
has to be chosen in front of the square root in Eq.\ (\ref{fe33b}).
The discontinuity of the order parameter on the transition line, Eq.\ (\ref{fe33}), is given by
\begin{eqnarray}
   \rho(T_1,P_1) = - \frac{B(T_1)}{2C'(T_1)} ,                 % (36)
\label{fe33a}
\end{eqnarray}
where we remind that $T_1$ is an implicit function of $P_1$.
With the parameters of Table \ref{table1} we obtain $\rho_1=-0.0425$.
Here and in the following we assume that the quantities $B(T_1)$, $C'(T_1)$ and $x^{(2)}(T_1)$,
all which are single expectation values, are not affected by moderate variations of pressure near $P_1$.
In the ordered phase we obtain by means of Eqs.\ (\ref{fe33b}) and (\ref{fe33}):
\begin{eqnarray}
   & & \rho(T_1,P^>) = \nonumber \\
   & & \frac{-3B(T_1) - \sqrt{B^2(T_1) + 192a |\Lambda| \kappa_L C'(T_1) \triangle P}}{8C'(T_1)} , \nonumber \\   % (37)
\label{fee37}
\end{eqnarray}
where $\triangle P \ge 0$.
Since $\rho(T_1,P_1) < 0$, the order parameter amplitude increases in absolute value with the increment of
pressure above $P_1$. In Fig.\ \ref{fig1} we have calculated the order parameter amplitude by means of Eq.\ (\ref{fee37}).
Here and below the cubic lattice constant $a$ stands for $a_{\alpha}$ in the ordered phase and
$a_{\gamma}$ in the disordered phase.
%%%%%%%%%%%%%%%%%%%%%%%%%%%%%%%%%%%%%%%%%%%%%%%%%%%%%%%%%%%%%%%%%%
%
%------------------------------------------------------
%    FIGURE 2
%------------------------------------------------------
%
%%%%%%%%%%%%%%%%%%%%%%%%%%%%%%%%%%%%%%%%%%%%%%%%%%%%%%%%%%%%%%%%%

In the following we will need the wave number dependent order parameter susceptibilities
$\chi_{\alpha \alpha}^{\rho \rho}(\vec{k})$ in the disordered phase and in the ordered phase.
The susceptibility is defined in terms of the fluctuations $\delta \rho^{(\alpha)}(\vec{k})$ of the local
order parameter from its average value:
\begin{eqnarray}
   \delta \rho^{(\alpha)} (\vec{k}) = \rho_Q^{(\alpha)}(\vec{k}) - \sqrt{N}\; \rho \; \delta_{\vec{k},\vec{k}_i^X} ,  %(38)
\label{fe38n}
\end{eqnarray}
where $i=x,y,z$ for $\alpha=3,1,2$, respectively.
Hence,
\begin{eqnarray}
   \chi_{\alpha \alpha}^{\rho \rho}(\vec{k}) = \frac{1}{T} \langle \delta \rho^{(\alpha)}(\vec{k}) \delta \rho^{(\alpha)}(-\vec{k}) \rangle .  %(39)
\label{fe39n}
\end{eqnarray}

We first consider the disordered phase.
In case of a rigid lattice we obtain within molecular field theory
\begin{eqnarray}
   \chi_{\alpha \alpha}^{\rho \rho}(\vec{k}) & \equiv & \frac{1}{T} \langle \rho_Q^{(\alpha)}(\vec{k}) \rho_Q^{(\alpha)}(-\vec{k}) \rangle
   \nonumber \\
   & = & x^{(2)} \left[ 1 \cdot T + x^{(2)} v(\vec{k}) \right]^{-1}_{\alpha \alpha} .
\label{fe34}
\end{eqnarray} %35
We recall that $v(\vec{k})$ is the quadrupolar interaction matrix Eq.\ (\ref{m10}) while $1$ is the
$3 \times 3$ unit matrix.
The single particle expectation value $x^{(2)} \equiv \langle \rho^{(\alpha)}(\vec{n}) \rho^{(\alpha)}(\vec{n}) \rangle$
has the same value for $\alpha=1,2$, and 3.
Since we consider order parameter fluctuations near the phase transition, only wave vectors $\vec{k}$ near the star of
$\vec{k}^X$ are relevant.
In accordance with the condensation scheme (\ref{m11a}), (\ref{m11b}) the matrix $v(\vec{k})$ becomes diagonal at
$\vec{k}=\vec{k}^X_z+\delta \vec{k}$, where $|\delta \vec{k}| < |\vec{k}^X_z|$.
Expansion about $\vec{k}^X_z$ gives
\begin{eqnarray}
   v_{2 2}(\vec{k}) = -4 \gamma + \frac{\gamma a^2}{2} (\delta k_z^2 + \delta k_x^2) .
\label{fe35}
\end{eqnarray}
Similar expressions with cyclic permutation of indices are obtained for $v_{1 1}(\vec{k})$ and $v_{3 3}(\vec{k})$
near $\vec{k}^X_y$ and $\vec{k}^X_x$, respectively.
In case of a non-rigid lattice we see from Eq.\ (\ref{fe32b}) that
the applied pressure acts as addition to the quadrupolar interaction at $\vec{k}^X$.
Replacing in Eq.\ (\ref{fe35}) $4 \gamma$ by $4 \tilde{\gamma}$, we rewrite for the order parameter
susceptibility in the disordered phase at temperature $T_1$ and pressure $P^<$:
\begin{eqnarray}
   \chi_{2 2}^{\rho \rho}(\vec{k}) = \frac{2\, \xi_d^2(T_1,P^<) }{a^2 \gamma [1 + \xi_d^2(T_1,P^<)(\delta k_z^2 + \delta k_x^2)]} .   % (42)
\label{fee40}
\end{eqnarray}
Here the correlation length $\xi_d$ is defined by
\begin{eqnarray}
   \xi_d(T_1,P^<) = \left[ \frac{x^{(2)}(T_1) \gamma a^2}{2( T_1-4\tilde{\gamma}(P^<)\, x^{(2)}(T_1)) } \right]^{1/2} .    % (43)
\label{fee41}
\end{eqnarray}
Expressions similar to Eq.\ (\ref{fee40}) are obtained for $\chi_{1 1}^{\rho \rho}(\vec{k})$ and
$\chi_{3 3}^{\rho \rho}(\vec{k})$ near $\vec{k}=\vec{k}^X_y$ and $\vec{k}^X_x$, respectively, and
with $(\delta k_z^2 + \delta k_x^2)$ replaced by $(\delta k_y^2 + \delta k_z^2)$ and $(\delta k_x^2 + \delta k_y^2)$, respectively.

Making use of Eqs.\ (\ref{fe33}) and (\ref{fe32b}), we obtain the correlation length
\begin{eqnarray}
   \xi_d(T_1,P^<) = \left[ \frac{3 \gamma a^2 C'(T_1) }{B^2(T_1) + 24C'(T_1) a|\Lambda|\kappa_L \triangle P]} \right]^{1/2} ,   % (44)
   \nonumber \\
\label{fee42}
\end{eqnarray}
where $\triangle P < 0$.
With increasing pressure $P^< \rightarrow P_1$ in the disordered phase, the correlation length reaches its maximum value
\begin{eqnarray}
   \xi_d(T_1,P_1) = \left[ \frac{3 \gamma a^2 C'(T_1)}{B^2(T_1)} \right]^{1/2} .       % Eq. (45)
\label{fee43}
\end{eqnarray}
toward the onset of the phase transition.
Likewise the susceptibility becomes maximum at $\vec{k}=\vec{k}_x^X$ and $(T_1,P_1)$:
\begin{eqnarray}
   \chi_{2 2}^{\rho \rho}(\vec{k}_x^X) = \frac{2\, \xi_d^2(T_1,P_1) }{a^2 \gamma} .      % (46)
\label{fee43a}
\end{eqnarray}

\begin{figure}
\resizebox{0.45\textwidth}{!} {
\includegraphics{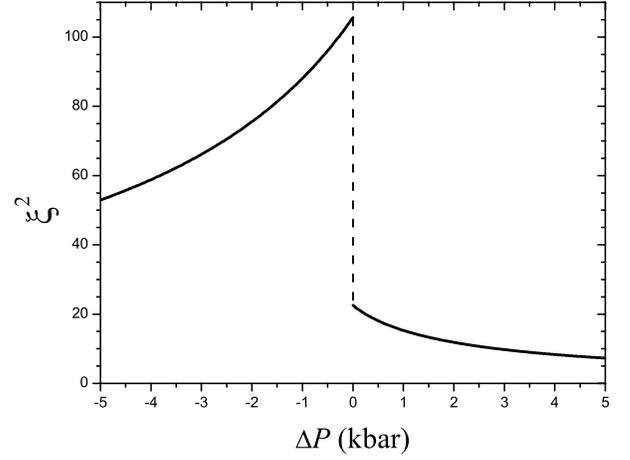}
}

\caption{ Pressure dependence of the squared correlation length $\xi^2(T_1,P)$ (in units {\AA}$^2$), $P_1$ taken as origin.
}
\label{fig2}

\end{figure}

In the ordered phase the order parameter susceptibility is given by
\begin{eqnarray}
   \chi_{\alpha \alpha}^{\rho \rho}(\vec{k}) \equiv \frac{1}{T} \left[ \langle \rho_Q^{(\alpha)}(\vec{k}) \rho_Q^{(\alpha)}(-\vec{k})    % (47)
    \rangle \,-\, N \rho^2 \; \delta_{\vec{k},\vec{k}_i^X} \right] .
\label{fe40}
\end{eqnarray}
Applying mean-field theory \cite{Cha} we obtain for the quadrupolar model on a rigid lattice:
\begin{eqnarray}
   \chi_{\alpha \alpha}^{\rho \rho}(\vec{k})
   = x^{(2)} \left[ 1 \cdot T + x^{(2)} v(\vec{k}) + 1 \cdot x^{(2)} D(T) \rho^2 \right]^{-1}_{\alpha \alpha} ,   % (48)
   \nonumber \\
\label{fe41}
\end{eqnarray}
where
\begin{eqnarray}
 & &D(T) = {\frac{T}{2(x^{(2)})^4}} \nonumber \\
 & & \times \left[ \,5(x^{(2)})^2 - x_{1111}^{(4)} - 2x_{1122}^{(4)}     % (49)
 + {\frac{8(x_{123}^{(3)})^2}{x^{(2)}}} \right] . \quad
   \label{fe42}
\end{eqnarray}
Numerical calculation show that $D(T) > 0$.
In expression (\ref{fe40}) $\rho$ is the order parameter amplitude.
Proceeding now as before we consider a nonrigid lattice and an external pressure.
Using Eqs.\ (\ref{fe32b}) and (\ref{fe35}) we define the correlation length in the ordered phase
at temperature $T_1$ and pressure $P^>$:
\begin{eqnarray}
   & & \xi_Q(T_1,P^>) =    \nonumber \\
   & & \left[ \frac{x^{(2)}(T_1) \gamma a^2 }{2(T_1-4\tilde{\gamma}(P^>) x^{(2)}(T_1)+ x^{(2)}(T_1) D(T_1)\rho^2(T_1,P^>))} \right]^{1/2} .  % (50)
   \nonumber \\
   \label{fe43}
\end{eqnarray}
Here the subscript $Q$ stands for ordered quadrupolar phase.
The corresponding order parameter susceptibility for the component $\alpha=2$ is then given by
\begin{eqnarray}
   \chi_{2 2}^{\rho \rho}(\vec{k}) = \frac{2\, \xi_Q^2(T_1,P^>)}{a^2 \gamma [1 + \xi_Q^2(T_1,P^>)(\delta k_z^2 + \delta k_x^2)]} .   % (51)
\label{fe44}
\end{eqnarray}
Notice that the discontinuity of the order parameter at the first order phase transition
leads to a drop of the correlation length and hence of the order parameter susceptibility.
With $\rho(T_1,P_1)$ given by Eq.\ (\ref{fe33a}), we obtain
\begin{eqnarray}
   \xi_Q(T_1,P_1) = \left[ \frac{3 \gamma\, a^2\, C'(T_1)}{B^2(T_1)[1 + 3 D(T_1)/ 2 C'(T_1)]} \right]^{1/2} ,   % Eq. (52)
   \nonumber \\
\label{fee49}
\end{eqnarray}
and since $D/C' > 0$, $\xi_Q(T_1,P_1) < \xi_d(T_1,P_1)$.
The correlation length in the ordered phase is calculated by means of Eqs.\ (\ref{fe43}), (\ref{fee37}) and (\ref{fe33}).
The result reads
\begin{eqnarray}
   \xi_Q(T_1,P^>) & = & ( 3 \gamma a^2 C'(T_1) )^{1/2}       \nonumber \\
    & & \left[ B^2(T_1) + 6 C'(T_1) D(T_1) \rho^2(T_1,P_1)   \right.   \nonumber \\                     % (53)
    & & \left. + 24 a\, \alpha(T_1) |\Lambda|\kappa_L \triangle P \right]^{-1/2}
\label{fee50}
\end{eqnarray}
where
\begin{eqnarray}
   \alpha(T_1) = 3D(T_1) - C'(T_1) .    % (54)
\label{fee51}
\end{eqnarray}
Numerical evaluation of the quadrupolar model shows that $\alpha(T_1) > 0$.
Hence the correlation length decreases with increasing pressure in the ordered phase.
Using Eq.\ (\ref{fee42}) for $\xi_d(T_1,P^<)$ and Eq.\ (\ref{fe43}) for $\xi_Q(T_1,P^>)$
we have calculated the squared correlation length $\xi^2_{d(Q)}(T_1,P^{>(<)})$
as a function of pressure. The plot is shown in Fig.\ \ref{fig2}. From Eqs.\ (\ref{fee40}) and (\ref{fe44})
for $\chi_{2 2}^{\rho \rho}(\vec{k})$ we see that Fig.\ \ref{fig2} also describes the pressure dependence of
the order parameter susceptibility at $\vec{k}=\vec{k}_x^X$.
%%%%%%%%%%%%%%%%%%%%%%%%%%%%%%%%%%%%%%%%%%%%%%%%%%%%%%%%%%%%%%%%%%
%
%------------------------------------------------------
%    FIGURE 3
%------------------------------------------------------
%
%%%%%%%%%%%%%%%%%%%%%%%%%%%%%%%%%%%%%%%%%%%%%%%%%%%%%%%%%%%%%%%%%

The expressions of the order parameter susceptibility will be used in the next section where we study the
temperature and pressure dependence of the elastic response near the first order phase transition.

%%%%%%%%%%%%%%%%%%%%%%%%%%%%%%%%%%%%%%%%%%%%%%%%%%%%%%%%%%%%%%%%%%%%%%%%%%%%%%%%%%%%
% 4
\section{Elastic response}
\label{sec:el_res}

We will derive the static displacement -- displacement response function matrix \cite{Got}
by using the well known Green's functions techniques \cite{Zub}.
Inversion of the response function in the long wavelength limit leads to the elastic constants.

The Fourier transform of the retarded Green's function of two operators $A$ and $B$ is defined by
\begin{eqnarray}
    \langle \langle A;B \rangle \rangle = -i \int_{-\infty}^{+\infty} dt e^{izt} \Theta(t) \langle [A(t),B(0)] \rangle ,
\label{m22}
\end{eqnarray}
with frequency $z=\omega+i\epsilon$, $\epsilon \rightarrow 0^{+}$.
The skew brackets $\langle ... \rangle$ stand for a thermal average with the system's Hamiltonian $H$.
In Heisenberg representation the time dependence reads $A(t)=e^{i {\cal H}t} A e^{-i {\cal H}t}$, ${\cal H}=H/\hbar$.
We quote the equation of motion \cite{Zub}:
\begin{eqnarray}
   z \langle \langle A;B \rangle \rangle_z = \langle [A,B] \rangle + \langle \langle [A,{\cal H}]; B \rangle \rangle_z ,
\label{m23}
\end{eqnarray}
with the identity
\begin{eqnarray}
    \langle \langle [A,{\cal H}]; B \rangle \rangle_z = - \langle \langle A; [B,{\cal H}] \rangle \rangle_z .
\label{m24}
\end{eqnarray}
The dynamic displacement -- displacement Green's function is defined by
\begin{eqnarray}
    D_{ij}(\vec{q},z) = \langle \langle u_i(\vec{q}); u_j(-\vec{q}) \rangle \rangle_z ,
\label{m25}
\end{eqnarray}
with lattice displacements and conjugate momenta specified in Eqs.\ (\ref{m14a}), (\ref{m14b}).
We recall that the Hamiltonian is given by Eq.\ (\ref{m1}), where the parts have been specified subsequently in Sect.\ \ref{sec:model}.
In the following we treat $\Omega^{(i)}(\vec{q})$ as a dynamic variable.
Thereby we retain only the fluctuation contribution of the order parameter variable, writing
\begin{eqnarray}
   \Omega^{(i)}(\vec{q}) = \sum_{\vec{k}} {\sum_{\alpha}}' \delta \rho^{(\alpha)}(-\vec{k}+\vec{q})\; \delta \rho^{(\alpha)}(\vec{k}) .
\label{m59n}
\end{eqnarray}
Applying twice the equation of motion (\ref{m23}) to $D_{ij}(\vec{q},z)$ we use the commutation rules Eq.\ (\ref{m15})
as well as the fact that the electronic variables $\rho^{(\alpha)}$ and $\Omega^{(i)}$ commute with the lattice variables $\vec{u}$ and $\vec{p}$.
The result reads
\begin{eqnarray}
    & & (z^2 \delta_{ik}-M_{ik}(\vec{q})) D_{kj}(\vec{q},z) = \nonumber \\
    & & \hbar \delta_{ij} + \frac{i\Lambda a}{\sqrt{Nm}} q_i
    \langle \langle \Omega^{(i)}(\vec{q}); u_j(-\vec{q}) \rangle \rangle_z .
\label{m26}
\end{eqnarray}
Here the summation is understood over the repeated index $k$.
We recall that $\Omega^{(i)}(\vec{q})$ has been defined by Eq.\ (\ref{m20}).
In a similar way we obtain by using Eq.\ (\ref{m24})
\begin{eqnarray}
    (z^2 \delta_{ik}-M_{ik}(\vec{q})) \langle \langle \Omega^{(i)}(\vec{q}); u_k(-\vec{q}) \rangle \rangle_z = \nonumber \\
   -\frac{i\Lambda a}{\sqrt{Nm}} q_j
    \langle \langle \Omega^{(i)}(\vec{q}); \Omega^{(j)}(-\vec{q}) \rangle \rangle_z .
\label{m27}
\end{eqnarray}
Taking the static limit $z=0$ in Eqs.\ (\ref{m26}) and (\ref{m27}) and combining the results we find
\begin{eqnarray}
  & & D_{ij}(\vec{q},z=0) = M_{ir}^{-1}(\vec{q}) \nonumber \\
  & & \times \left[ -\hbar \delta_{r j} + \frac{\Lambda^2 a^2}{Nm} q_r q_s M_{j s}^{-1}(\vec{q})
    \langle \langle \Omega^{(r)}(\vec{q}); \Omega^{(s)}(-\vec{q}) \rangle \rangle_0 \right] . \nonumber \\
\label{m28}
\end{eqnarray}
Here the subscript $0$ stands for $z=0$.

We introduce the static susceptibilities \cite{Got}
\begin{eqnarray}
   & & \chi_{ij}^{uu}(\vec{q})= -\frac{1}{\hbar} D_{ij}(\vec{q},z=0)  ,
\label{m29} \\
   & & \chi_{r s}^{\Omega \Omega}(\vec{q})= -\frac{1}{\hbar} \langle \langle \Omega^{(r)}(\vec{q}); \Omega^{(s)}(-\vec{q}) \rangle \rangle_0 ,
\label{m30}
\end{eqnarray}
and rewrite Eq.\ (\ref{m28}) as
\begin{eqnarray}
 \chi_{ij}^{uu}(\vec{q}) = M_{ir}^{-1}(\vec{q}) \left[ \delta_{r j} + \frac{\Lambda^2 a^2}{Nm} q_r q_s M_{j s}^{-1}(\vec{q}) \,
   \chi_{r s}^{\Omega \Omega}(\vec{q}) \right] . \nonumber \\
\label{m31}
\end{eqnarray}
We remind that in the long wavelength limit the static displacement susceptibility is related to the elastic constants \cite{Got} $C_{ijkl}$ by
\begin{eqnarray}
 \left( \chi^{uu}(\vec{q}) \right)^{-1}_{ik} = \mu^{-1}\, C_{ijkl}\, q_j \, q_l ,
\label{m32}
\end{eqnarray}
where $\mu=4m/a^3$ is the mass density.
Hence Eq.\ (\ref{m31}) allows us to calculate the elastic constants in presence of the coupling $H^{(e,L)}$ between
electronic and lattice degrees of freedom.
We recall that in absence of the coupling the bare elastic constants are given by $M_{ik}(\vec{q})=\mu^{-1} q_j q_l C^0_{ijkl}$ [see Eq.\ (\ref{m16})].
For cubic crystals we have in Voigt's notation the elastic constants $C_{xxxx} \equiv c_{11}$, $C_{xyxy} \equiv c_{44}$, $C_{xxyy} \equiv c_{12}$.
With $\vec{q}=\vec{q}_1 \equiv q_x(1,0,0)$ and $i=j=x$, Eq.\ (\ref{m31}) reduces in the long wavelength limit to
\begin{eqnarray}
 c_{11} = c_{11}^0 \left[ 1 + 4 \frac{\Lambda^2}{a c_{11}^0 N}
   \chi_{xx}^{\Omega \Omega}(\vec{q}_1=0) \right]^{-1} .                            %(67)
\label{m33}
\end{eqnarray}
Here we have made use of Eqs.\ (\ref{m32}) and (\ref{m16}).
On the basis of the Hamiltonian, Eq.\ (\ref{m1}), the obtained expressions for $\chi_{ij}^{uu}(\vec{q})$ and $c_{11}$ are
rigorous results.
Since $\chi_{xx}^{\Omega \Omega}(\vec{q}_1=0) \ge 0$, the quadrupolar-elastic coupling, Eq.\ (\ref{m19}), leads to a reduction of
the elastic constant $c_{11}$ in comparison with the bare quantity $c_{11}^0$.
This reduction as a consequence of the $H^{(e,L)}$ coupling is responsible for the relative softening of the corresponding
$L$[001] phonon branch in $\gamma$-Ce \cite{Sta,Kri}.
In contradistinction to $c_{11}$, the shear elastic constant $c_{44}$ is not affected by the coupling of the lattice
to quadrupolar electronic density fluctuations.
As we have shown in Sec.\ II, there is no coupling to shear strains $\epsilon_{xy}$ in $H^{(e,L)}$, Eq.\ (\ref{m19}).
Hence taking $\Lambda=0$ in Eq.\ (\ref{m31}), we obtain for $\vec{q}=\vec{q}_1$ and $i=j=y$:
\begin{eqnarray}
   \left( \chi_{y y}^{u u}(\vec{q}_1) \right)^{-1} = M_{y y}(\vec{q}_1) ,               %(68)
\label{m34}
\end{eqnarray}
or equivalently by means of Eqs.\ (\ref{m16}) and (\ref{m32}) $c_{44}=c_{44}^0$.
Pressure experiments exhibit a discontinuity of the shear modulus \cite{Dec1} at the phase transition.
However, this effect is solely due to the lattice contraction and a concomitant change of interatomic forces.
Due to the absence of a direct coupling between lattice shears and orientational density fluctuations
there are no precursor effects.

We want to study the anomalous behavior of the elastic constant $c_{11}$ near the quadrupolar ordering transition.
From Eq.\ (\ref{m33}) it follows that the important quantity is the four-point (four factors $\rho$)
susceptibility $\chi_{xx}^{\Omega \Omega}(\vec{q}_1=0)$.
At high temperature and in the long wavelength regime such that the length scale of order parameter fluctuations is large
in comparison with the interatomic spacing we use classical statistical mechanics.
We then have
\begin{eqnarray}
   \chi_{xx}^{\Omega \Omega}(\vec{q}_1) = \frac{1}{T} \langle \delta \Omega^{(x)}(\vec{q}_1) \, \delta \Omega^{(x)}(-\vec{q}_1) \rangle , %(69)
\label{m35}
\end{eqnarray}
where
\begin{eqnarray}
   \delta \Omega^{(x)} (\vec{q}_1) = \Omega^{(x)}(\vec{q}_1) - \langle \Omega^{(x)}(\vec{q}_1) \rangle .  %(70)
\label{m70n}
\end{eqnarray}
In Appendix we show that in the limit $\vec{q}_1 \rightarrow 0$
\begin{eqnarray}
   \chi_{xx}^{\Omega \Omega}(\vec{q}_1=0) = 2T \sum_{\vec{k}} {\sum_{\alpha}}' ( \chi_{\alpha \alpha}^{\rho \rho}(\vec{k}) )^2 , %(71)
\label{m36}
\end{eqnarray}
with $\alpha=2$, 3. Here $\chi_{\alpha \alpha}^{\rho \rho}(\vec{k})$ is the order parameter susceptibility for the
component $\alpha$.
Cubic symmetry implies that
\begin{eqnarray}
   \sum_{\vec{k}} \left( \chi_{2 2}^{\rho \rho}(\vec{k}) \right)^2 = \sum_{\vec{k}} \left( \chi_{3 3}^{\rho \rho}(\vec{k}) \right)^2 . %(72)
\label{m38}
\end{eqnarray}

We calculate the right hand side of Eq.\ (\ref{m36}) by using
\begin{eqnarray}
   \sum_{\vec{k}} (\chi_{2 2}^{\rho \rho}(\vec{k}) )^2 = \frac{V}{(2\pi)^3} \int d^3k (\chi_{2 2}^{\rho \rho}(\vec{k}) )^2 ,  %(73)
\label{m45}
\end{eqnarray}
where $V=Na^3/4$ is the volume of the crystal and $N$ the number of atoms.
So far the considerations of the present section are general and hold for the disordered as well as for the ordered phase.
Since the expressions of the order parameter susceptibility in the disordered and in the ordered phase, Eqs.\ (\ref{fee40}) and (\ref{fe44}),
respectively, exhibit the same wave vector dependence, we can treat both cases simultaneously.
Since the integrand in Eq.\ (\ref{m45}) does not depend on $k_y$, the integration over $d k_y$ in the
interval $\pm 2\pi/a$ yields $4\pi/a$.
On the other hand the integrand vanishes for large values of $|\delta k_z^2 + \delta k_x^2|$.
Hence we integrate over a circle and extend the radius to $\infty$.
Taking into account Eq.\ (\ref{m38}), we obtain
\begin{eqnarray}
   \chi_{xx}^{\Omega \Omega}(\vec{q}=0) = \frac{2 T N \xi_{\nu}^2}{\pi a^2 \gamma^2} ,   %(74)
\label{m57}
\end{eqnarray}
where $\xi_{\nu}$ stands for $\xi_{d}$ in the disordered phase and for $\xi_{Q}$ in the ordered quadrupolar phase.
Substitution of the result into Eq.\ (\ref{m33}) gives the longitudinal elastic constant
\begin{eqnarray}
 c_{11} = c_{11}^0 \left[ 1 + \Xi_{\nu}(T_1,P) \right]^{-1} ,   %(75)
\label{m58}
\end{eqnarray}
where we have defined
\begin{eqnarray}
 \Xi_{\nu}(T_1,P) = \frac{8 \Lambda^2 T \xi_{\nu}^2}{c_{11}^0 \pi a^3 \gamma^2 }  ,   %(76)
\label{m76n}
\end{eqnarray}
with $\xi_{\nu}=\xi_{d}$, Eq.\ (\ref{fee42}), in the disordered phase and $\xi_{\nu}=\xi_{Q}$, Eq.\ (\ref{fee50}), in the ordered
phase.

Near the phase transition in the disordered phase we have
\begin{eqnarray}
 \left. c_{11}(T_1,P_<) \right|_d = c_{11}^0 \left[ 1 + \Xi_{d}(T_1,P^<) \right]^{-1} .   %(77)
\label{m73}
\end{eqnarray}
The increase of $\xi_d(T_1,P^<)$ or equivalently of $\Xi_{d}(T_1,P^<)$ with $\triangle P < 0$ leads to a decrease of $c_{11}$ which reaches its minimum value
\begin{eqnarray}
 \left. c_{11}(T_1,P_1) \right|_d = c_{11}^0 \left[ 1 + \Xi_{d}(T_1,P_1) \right]^{-1}    %(78)
\label{m74}
\end{eqnarray}
at $P^< = P_1$.

On the other hand at the onset of the phase transition one has
\begin{eqnarray}
 \left. c_{11}(T_1,P_1) \right|_Q = c_{11}^0 \left[ 1 + \Xi_{Q}(T_1,P_1) \right]^{-1} .    %(79)
\label{m75}
\end{eqnarray}
The discontinuity of the order parameter and the concomitant drop of the correlation length $\xi_Q(T_1,P_1) < \xi_d(T_1,P_1)$
results in a positive jump of $c_{11}$ at the first order phase transition:
\begin{eqnarray}
 & & \left. c_{11}(T_1,P_1) \right|_Q - \left. c_{11}(T_1,P_1) \right|_d =    \nonumber \\
 & & \frac{8 \Lambda^2 T_1}{\pi a^3 \gamma^2}
 \frac{[ \xi_d^2(T_1,P_1) - \xi_Q^2(T_1,P_1) ]}{[1 + \Xi_{d}(T_1,P_1) ][ 1 + \Xi_{Q}(T_1,P_1) ]} .                            %(80)
\label{m80n}
\end{eqnarray}
Here we have approximated $a$ in the prefactor by $(a_{\alpha}+a_{\gamma})/2$.

In the ordred phase the decrease of the correlation length $\xi_Q(T_1,P^>)$ with increasing pressure leads to
an increase of
\begin{eqnarray}
 \left. c_{11}(T_1,P^>) \right|_Q = c_{11}^0 \left[ 1 + \Xi_{Q}(T_1,P^>) \right]^{-1} .    %(81)
\label{m76}
\end{eqnarray}
The scenario described by Eqs.\ (\ref{m73})--(\ref{m76}) is in full agreement with experiment.
Indeed the pioneering experiments on elastic properties of Ce under pressure by Voronov et al.\ \cite{Vor}
as well as recent high resolution ultrasonic measurements \cite{Dec1} show that the
propagation velocity of longitudinal ultrasonic waves decreases when the phase transition toward the $\alpha$ phase
is approached with increasing pressure in the $\gamma$ phase.
At the first order phase transition the longitudinal sound velocity exhibits a stepwise increase.
With increasing pressure in the $\alpha$ phase the longitudinal sound velocity increases continuously.
A corresponding behavior of the bulk modulus as a function of pressure was deduced from high-resolution neutron and
synchrotron x-ray powder diffraction \cite{Jeo} experiments.
It was argued that the softening of the bulk modulus in the $\gamma$ phase with increasing pressure $P \rightarrow P_1$
is a direct consequence of the softening of $c_{11}$.

We close this section by observing that in C$_{60}$ fullerite there occurs a marked lowering of the bulk modulus $B$ if
at fixed $T$ the fcc$\rightarrow$sc phase transition is approached with increasing pressure \cite{Lun}.
The phenomenon is attributed to orientational reordering of the C$_{60}$ molecules.
Given the similarities of the lattice related phenomena at the $\gamma \rightarrow \alpha$ phase transition in Ce and the
orientational phase transition in C$_{60}$ fullerite, we conclude that these transitions are isomorphic.
From the mathematical point of view \cite{Princ} all essential aspects ($Fm{\bar 3}m \rightarrow Pa{\bar 3}$) are the same,
while the constituents (electronic quadrupolar densities versus icosahedral molecules) are different.
%%%%%%%%%%%%%%%%%%%%%%%%%%%%%%%%%%%%%%%%%%%%%%%%%%%%%%%%%%%%%%%%%%
%
%------------------------------------------------------
%    FIGURE 4
%------------------------------------------------------
\begin{figure}
\resizebox{0.45\textwidth}{!} {
\includegraphics{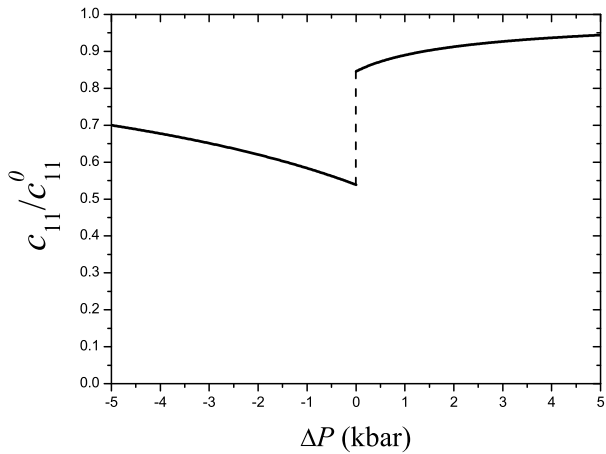}
}

\caption{ Pressure dependence of $c_{11}$ ($\Lambda=5\Lambda_0$), $P_1$ taken as origin.
}
\label{fig3}

\end{figure}
%
%%%%%%%%%%%%%%%%%%%%%%%%%%%%%%%%%%%%%%%%%%%%%%%%%%%%%%%%%%%%%%%%%

%%%%%%%%%%%%%%%%%%%%%%%%%%%%%%%%%%%%%%%%%%%%%%%%%%%%%%%%%%%%%%%%%%%%%%%%%
\section{Conclusions}
\label{sec:con}

The elastic properties of a model of interacting 4f-electron quadrupolar densities
on a compressible fcc lattice have been investigated as a function of temperature and pressure.
On the basis of previous theoretical work \cite{NM1a,NM1b,NM2} this model,
supported by recent nuclear spectroscopy experiments \cite{Tsv}, is taken
as representative for the $\gamma-\alpha$ ``isostructural" phase transition in Ce.
In particular, we have studied by analytical theory the pressure dependent anomalies of
the elastic constant $c_{11}$ at the phase transition from the quadrupolar orientationally
disoredered phase which we identify with the $\gamma$-phase (space group $Fm{\bar 3}m$)
to the quadrupolar orientationally ordered phase which we identify with the $\alpha$-phase (space group $Pa{\bar 3}$).
As a result we find that $c_{11}$ (equivalently the longitudinal sound velocity) decreases
by approaching the phase transition from the disoredered phase with increasing pressure $P^< \rightarrow P_1$,
at the first order phase transition $c_{11}$ exhibits a positive jump,
in the ordered phase $c_{11}$ increases continuously with pressure $P^> > P_1$.
On the other hand we find that the elastic constant $c_{44}$ (equivalently the shear sound velocity)
exhibits no precursor effects near the transition.
These theoretical results are in full qualitative agreement with experiments [34,21,22,33] on elastic anomalies at the
$\gamma-\alpha$ ``isostructural" phase transition in solid Ce.

We notice that the Hamiltonian $H^{(e,L)}$, Eq.\ (\ref{m17}), which is quadratic in the electronic quadrupolar
order parameter variables and linear in the lattice displacements, accounts as well for the elastic anomalies as for
the ``isostructural" lattice contraction.
The Hamiltonian is reminiscent from the compressible Ising model \cite{Wag,Ber}
which is quadratic in the spin variables and linear in the lattice displacements.
However, in the present case symmetry properties are more subtle and account for the interplay between
antiferroquadrupolar order in the electronic densities and ``isostructural" lattice contraction.

Here a remark on symmetry reduction at the phase transition is in order.
The quadrupolar interaction has its origin in the repulsive Coulomb interaction between 4f-electrons
on neighboring atoms. The ordering of the electronic quadrupoles on four sc sublattices
reduces the repulsion and acts as an effective attraction.
In reciprocal space the quadrupolar interaction matrix $v(\vec{q})$, Eq.\ (\ref{m10}),
has negative eigenvalues at the $X$-point of the Brillouin zone of the fcc lattice.

Phonon dispersions of Ce measured by synchrotron radiation show pressure dependent anomalies related to
the $\gamma-\alpha$ transition at the $X$-point of the Brillouin zone \cite{Kri}.
The explanation of these experiments is an outstanding challenge for theoretical work.

A further problem is the quantitative improvement of the theory which should lead to increase the magnitude
of the coupling coefficient $\Lambda$ in Eq.\ (\ref{m19}) for $H^{(e,L)}$.
At present we believe that the theory should be extended by including the intrasite coupling between
4f-electrons and conduction electrons (say 5d) and a possible influence of this coupling on the
interatomic bonding.
An indication of the relevance of such a mechanism is provided by recent work on the Ce dimer \cite{NikA}.
Likely such an extension of the theory would also contribute to elucidate the concept of a Mott transition \cite{Joh,Joh2,Gof}
versus a Kondo scenario \cite{All,Lav,Ful,Liu} at the $\gamma-\alpha$ transition.

\acknowledgements
The authors acknowledge useful discussions with K. Parlinski, M. Krisch, F. Decremps, R.M. Pick, G. Roth, G. G\"untherodt,
A.V. Tsvyashchenko and E.V. Tkalya.
Financial support has been provided by the research group Theory of Condensed Matter, University of Antwerp and by the
Institut f\"ur Kristallographie, RWTH Aachen.

\appendix
\section{}

Here we express $\chi_{xx}^{\Omega \Omega}(\vec{q}_1=0)$ as a function of the order
parameter susceptibility $\chi_{\alpha \alpha}^{\rho \rho}(\vec{k})$.
Starting from Eqs.\ (\ref{m35}) and (\ref{m70n}) we first notice that
\begin{eqnarray}
  \langle \Omega^{(x)}(\vec{q}_1) \rangle = \langle \Omega^{(x)}(\vec{q}_1=0) \rangle \; \delta_{\vec{q}_1,0} ,
\label{a1}
\end{eqnarray}
where
\begin{eqnarray}
  \langle \Omega^{(x)}(\vec{q}_1=0) \rangle  = {\sum_{\alpha} }' \sum_{\vec{k}} \langle \delta \rho^{(\alpha)}(\vec{k}) \, \delta \rho^{(\alpha)}(-\vec{k}) \rangle .
\label{a2}
\end{eqnarray}
We rewrite Eq.\ (\ref{m35}) as
\begin{eqnarray}
   \chi_{xx}^{\Omega \Omega}(\vec{q}_1) = \frac{1}{T} \left[ \langle \Omega^{(x)}(\vec{q}_1) \Omega^{(x)}(-\vec{q}_1) \rangle
   - \langle \Omega^{(x)}(\vec{q}_1=0) \rangle^2 \right] . \nonumber \\
\label{a3}
\end{eqnarray}
We then approximate the four-point function
\begin{eqnarray}
& & \langle \Omega^{(x)}(\vec{q}_1) \Omega^{(x)}(-\vec{q}_1) \rangle \equiv \sum_{\vec{k}} \sum_{\vec{p}} {\sum_{\alpha}}' {\sum_{\beta}}' \nonumber \\
& &   \langle \delta \rho^{(\alpha)}(-\vec{k}+\vec{q}_1)\, \delta \rho^{(\alpha)}(\vec{k})\, \delta \rho^{(\beta)}(-\vec{p}-\vec{q}_1)\,
  \delta \rho^{(\beta)}(\vec{p}) \rangle . \nonumber \\
\label{a4}
\end{eqnarray}
by the factorization scheme
\begin{eqnarray}
 & & \langle \Omega^{(x)}(\vec{q}_1) \Omega^{(x)}(-\vec{q}_1) \rangle \simeq     \nonumber \\
 & & \sum_{\vec{k}} \sum_{\vec{p}} {\sum_{\alpha}}' {\sum_{\beta}}'
  \langle \delta \rho^{(\alpha)}(-\vec{k})\, \delta \rho^{(\alpha)}(\vec{k})\rangle \langle
  \delta \rho^{(\beta)}(-\vec{p})\, \delta \rho^{(\beta)}(\vec{p}) \rangle    \nonumber \\
 & &+ \langle \delta \rho^{(\alpha)}(-\vec{k}+\vec{q}_1)\, \delta \rho^{(\beta)}(-\vec{p}-\vec{q}_1)\rangle \langle \delta \rho^{(\alpha)}(\vec{k})\, \delta \rho^{(\beta)}(\vec{p}) \rangle \; \nonumber \\
  & & \times \delta_{\vec{p}, -\vec{k}}\, \delta_{\alpha, \beta } \nonumber    \\
 & &+ \langle \delta \rho^{(\alpha)}(-\vec{k}+\vec{q}_1)\, \delta \rho^{(\beta)}(\vec{p})\rangle \langle \delta \rho^{(\alpha)}(\vec{k})\, \delta \rho^{(\beta)}(-\vec{p}-\vec{q}_1) \rangle \; \nonumber \\
 & & \times \delta_{\vec{p}, \vec{k}-\vec{q}_1}\, \delta_{\alpha, \beta } . \nonumber \\
 \label{a5}
\end{eqnarray}
or equivalently
\begin{eqnarray}
   & & \langle \Omega^{(x)}(\vec{q}_1) \Omega^{(x)}(-\vec{q}_1) \rangle = \langle \Omega^{(x)}(\vec{q}_1=0) \rangle^2  \nonumber \\
   & & + 2 \sum_{\vec{k}} {\sum_{\alpha}}'
    \langle \delta \rho^{(\alpha)}(-\vec{k}+\vec{q}_1)\, \delta \rho^{(\alpha)}(\vec{k}-\vec{q}_1) \rangle  \nonumber \\
   & & \times \langle \delta \rho^{(\alpha)}(\vec{k})\, \delta \rho^{(\alpha)}(-\vec{k}) \rangle .
\label{a6}
\end{eqnarray}
Substituting this result into Eq.\ (\ref{a3}) and using Eq.\ (\ref{fe39n}) we obtain
\begin{eqnarray}
   \chi_{xx}^{\Omega \Omega}(\vec{q}_1) = 2T \sum_{\vec{k}} {\sum_{\alpha}}'
   \chi_{\alpha \alpha}^{\rho \rho}(\vec{k}-\vec{q}_1) \, \chi_{\alpha \alpha}^{\rho \rho}(\vec{k}) .
\label{a7}
\end{eqnarray}
Taking the limit $\vec{q}_1=0$ leads to the result of Eq.\ (\ref{m36}).

%%%%%%%%%%%%%%%%%%%%%%%%%%%%%%%%%%%%%%%%%%%%%%%%%%%%%%%%%%%%%%%%%%%%%%%%%%%%%%%

%---------------- REFERENCES -------------------------------

\end{document}